\documentclass[12pt]{article}
\usepackage{graphicx}

\def\hybrid{\topmargin 0pt      \oddsidemargin 0pt
        \headheight 0pt \headsep 0pt
       \voffset-1cm
        \textwidth 6.25in       
       \textheight 9.5in       
        \marginparwidth 0.0in
        \parskip 5pt plus 1pt   \jot = 1.5ex}
\catcode`\@=11
\def\marginnote#1{}

\newcount\hour
\newcount\minute
\newtoks\amorpm
\hour=\time\divide\hour by60
\minute=\time{\multiply\hour by60 \global\advance\minute by-\hour}
\edef\standardtime{{\ifnum\hour<12 \global\amorpm={am}%
        \else\global\amorpm={pm}\advance\hour by-12 \fi
        \ifnum\hour=0 \hour=12 \fi
        \number\hour:\ifnum\minute<10 0\fi\number\minute\the\amorpm}}
\edef\militarytime{\number\hour:\ifnum\minute<10 0\fi\number\minute}

\def\draftlabel#1{{\@bsphack\if@filesw {\let\thepage\relax
   \xdef\@gtempa{\write\@auxout{\string
      \newlabel{#1}{{\@currentlabel}{\thepage}}}}}\@gtempa
   \if@nobreak \ifvmode\nobreak\fi\fi\fi\@esphack}
        \gdef\@eqnlabel{#1}}
\def\@eqnlabel{}
\def\@vacuum{}
\def\draftmarginnote#1{\marginpar{\raggedright\scriptsize\tt#1}}

\def\draftlabel#1{{\@bsphack\if@filesw {\let\thepage\relax
   \xdef\@gtempa{\write\@auxout{\string
      \newlabel{#1}{{\@currentlabel}{\thepage}}}}}\@gtempa
   \if@nobreak \ifvmode\nobreak\fi\fi\fi\@esphack}
        \gdef\@eqnlabel{#1}}
\def\@eqnlabel{}
\def\@vacuum{}
\def\draftmarginnote#1{\marginpar{\raggedright\scriptsize\tt#1}}

\def\draft{\oddsidemargin -.5truein
        \def\@oddfoot{\sl preliminary draft \hfil
        \rm\thepage\hfil\sl\today\quad\militarytime}
        \let\@evenfoot\@oddfoot \overfullrule 3pt
        \let\label=\draftlabel
        \let\marginnote=\draftmarginnote
   \def\@eqnnum{(\theequation)\rlap{\kern\marginparsep\tt\@eqnlabel}%
\global\let\@eqnlabel\@vacuum}  }

\def\numberbysection{\@addtoreset{equation}{section}
        \def\theequation{\thesection.\arabic{equation}}}

\def\underline#1{\relax\ifmmode\@@underline#1\else
        $\@@underline{\hbox{#1}}$\relax\fi}

\def\titlepage{\@restonecolfalse\if@twocolumn\@restonecoltrue\onecolumn
     \else \newpage \fi \thispagestyle{empty}\c@page\z@
        \def\thefootnote{\fnsymbol{footnote}} }

\def\endtitlepage{\if@restonecol\twocolumn \else  \fi
        \def\thefootnote{\arabic{footnote}}
        \setcounter{footnote}{0}}  
\relax

\hybrid

\newfont{\Bbb}{msbm10 scaled 1\@ptsize00}
\newfont{\Bbbb}{msbm7 scaled 1\@ptsize00}

\newcommand{\DDD}{\raise-1pt\hbox{$\mbox{\Bbbb D}$}}

\newcommand{\UUU}{\raise-1pt\hbox{$\mbox{\Bbbb U}$}}

\newcommand{\z}{\raise-1pt\hbox{$\mbox{\Bbbb Z}$}}

\def\beq{\begin{equation}}
\def\eeq{\end{equation}}
\def\p{\partial}

\newtheorem{lemma-definition}{Lemma-Definition}[section]

\newtheorem{proposition}{Proposition}[section]

\begin{document}

\begin{titlepage}

\title{Multi-variable reductions of the 
dispersionless DKP hierarchy}

\author{V.~Akhmedova\thanks{National Research University Higher School of 
Economics, 
20 Myasnitskaya Ulitsa, Moscow 101000, Russia,
e-mail: valeria-58@yandex.ru}
\and T.~Takebe 
\thanks{
National Research University Higher School of Economics,
20 Myasnitskaya Ulitsa,
Moscow 101000, Russia, e-mail: ttakebe@hse.ru}
\and A.~Zabrodin
\thanks{
National Research University Higher School of Economics,
20 Myasnitskaya Ulitsa,
Moscow 101000, Russia and
ITEP, 25
B.Cheremushkinskaya, Moscow 117218, Russia
e-mail: zabrodin@itep.ru}}

\date{July 2017}
\maketitle

\vspace{-7cm} \centerline{ \hfill ITEP-TH-16/17}\vspace{7cm}

\begin{abstract}

We consider multi-variable reductions of the dispersionless DKP
hierarchy (the dispersionless limit of the Pfaff lattice) in the 
elliptic parametrization.
The reduction is given by a system of elliptic
L\"owner equations 
supplemented by a system of partial differential equations of hydrodynamic type.
The compatibility conditions for the 
elliptic
L\"owner equations are derived. They 
are elliptic analogues of the Gibbons-Tsarev equations. 
We prove solvability of the hydrodynamic type system by means of the 
generalized hodograph method. The associated diagonal metric is proved to be of the
Egorov type.

\end{abstract}

\end{titlepage}

\vspace{5mm}

%



\section{Introduction}

The DKP hierarchy is one of the integrable hierarchies with 
$D_{\infty}$ symmetries introduced by 
M.Jimbo and T.Miwa in 1983 \cite{JimboMiwa}.
It was subsequently rediscovered and came to be
also known as the coupled KP hierarchy \cite{HO} and
the Pfaff lattice \cite{AHM,ASM}, see also
\cite{Kakei,IWS,Willox}. 
The solutions and the algebraic structure were studied in 
\cite{Kodama,Kodama1,AKM}, the relation to matrix integrals 
was elaborated in \cite{AHM,ASM,Kakei,Vandeleur,Orlov}.

The dispersionless version of the DKP hierarchy (the dDKP
hierarchy) was suggested in 
\cite{Takasaki07,Takasaki09}.
It is an infinite system
of differential equations for a real-valued function 
$F=F({\bf t})$ of the infinite number of 
(real) ``times''
${\bf t}=\{t_0, t_1, t_2, \ldots \}$.
The differential equations are obtained by expanding 
equations
\beq\label{D1}
e^{D(z)D(\zeta )F}\left (1-\frac{1}{z^2\zeta^2}\,
e^{2\p_{t_0}(2\p_{t_0} + D(z) + D(\zeta ))F}\right )=
1-\frac{\p_{t_1}D(z)F -\p_{t_1}D(\zeta )F}{z-\zeta},
\eeq
\beq\label{D2}
e^{-D(z)D(\zeta )F}
\, \frac{z^2 e^{-2\p_{t_0}D(z)F}-\zeta^2 e^{-2\p_{t_0}D(\zeta )F}}{z-\zeta}
=z+\zeta -\p_{t_1}\! \Bigl (2\p_{t_0} +D(z)+D(\zeta )\Bigr )F,
\eeq
where 
\beq\label{D3}
D(z)=\sum_{k\geq 1}\frac{z^{-k}}{k}\, \p_{t_k}
\eeq
in powers of $z$, $\zeta$.
The function $F$ corresponds to the logarithm of the tau function in the case of
the dispersionless KP hierarchy (cf., for example, \cite{tak-tak:95,take:14-1}).

In \cite{akh-zab:14-1,akh-zab:14-2} it was shown that equations (\ref{D1}), (\ref{D2}),
when rewritten in an elliptic pa\-ra\-met\-ri\-za\-tion in terms of
Jacobi's theta-functions $\theta_a(u,\tau )$, assume a nice 
and suggestive form which looks like a natural elliptic extension
of the dispersionless KP hierarchy:
\beq\label{DD4}
(z^{-1}-\zeta^{-1})e^{(\p_{t_0}+D(z))(\p_{t_0}+D(\zeta ))F}
=\frac{\theta_1(u(z)\! -\! 
u(\zeta ),\tau )}{\theta_4(u(z)\! -\! u(\zeta ),\tau )}\,.
\eeq
Here the function $u(z)$ is defined by
\beq\label{DD4a}
e^{\p_{t_0}(\p_{t_0}+D(z))F}=z \, 
\frac{\theta_1(u(z),\tau )}{\theta_4(u(z),\tau )}\,.
\eeq
The modular parameter $\tau$  
is a dynamical variable: 
$\tau = \tau ({\bf t})$. This feature suggests some 
similarities with the genus 1 Whitham equations \cite{Krichever}
and the integrable structures behind boundary value 
problems in plane
doubly-connected domains \cite{KMZ05}.
We assume that
$\tau$ is purely imaginary.

One may look for solutions of the hierarchy such that 
$u(z, {\bf t})$ and $\tau ({\bf t})$ depend on the times 
through a single variable $\lambda = \lambda ({\bf t})$:
$u(z, {\bf t})=u(z, \lambda ({\bf t}))$, 
$\tau ({\bf t})=\tau (\lambda ({\bf t}))$. 
In \cite{akh-zab:14-1} it was shown that
such one-variable reductions are classified by solutions of 
a differential equation which is an elliptic analogue of the 
famous L\"owner equation (see, e.g., 
\cite[Chapter 6]{Pommerenke}). In complex analysis, this 
``elliptic L\"owner equation'' is also known
as the Goluzin-Komatu equation \cite{Goluzin,Komatu}, 
see also \cite{Alexandrov,CDMG1,CDMG2,review}:
\beq\label{DD5}
\begin{array}{l}
4\pi {\rm i} \, \p_{\lambda}u(z, \lambda )=\Bigl [-\, \zeta_1 \Bigl ( u(z, \lambda ) \! +\! 
\xi (\lambda ), \, \frac{\tau}{2}\Bigr )+
\, \zeta_1 \Bigl ( 
\xi (\lambda ), \, \frac{\tau}{2}\Bigr )\Bigr ]\displaystyle{\frac{\p \tau}{\p \lambda}},
\end{array}
\eeq
where $\zeta_1 (u, \tau ):=\p_u \log \theta_1(u, \tau )$ and
$\xi (\lambda )$ is an arbitrary (continuous) function of $\lambda$ 
(the ``driving function'').
This equation is the basic element of the 
theory of parametric conformal maps from
doubly connected slit domains to annuli. A similar relation between the chordal L\"owner
equation and one-variable 
reductions of the dKP hierarchy was known since the seminal papers
by Gibbons and Tsarev \cite{GT1,GT2}. 
Further developments are discussed in
\cite{Manas1,Manas2,TT06,TTZ06,T13}.

In this paper we study diagonal $N$-variable reductions of the dDKP hierarchy
when $u$ depends on the times through $N$ real variables $\lambda_j$.
The starting point is the system of $N$ elliptic L\"owner equations which 
characterize the dependence of $u(z)$ on the variables $\lambda_j$:
\beq\label{DD5a}
\begin{array}{l}
4\pi {\rm i} \, \p_{\lambda_j}u(z, \{\lambda_i\} )=\Bigl [-\, \zeta_1 \Bigl ( u \! +\! 
\xi_j\, , \, \frac{\tau}{2}\Bigr )+
\, \zeta_1 \Bigl ( 
\xi_j\,  , \, \frac{\tau}{2}\Bigr )\Bigr ]\displaystyle{\frac{\p \tau}{\p \lambda_j}},
\end{array}
\eeq
Their compatibility condition is expressed as the elliptic Gibbons-Tsarev system
(see (\ref{GT1}), (\ref{GT2}) below).
The time dependence of the variables $\lambda_j$ is fixed by a system 
of quasi-linear partial differential equations of the form
\beq\label{intro1}
\frac{\p \lambda_j}{\p t_k}=\phi_{j,k}(\{\lambda_i\})\frac{\p \lambda_j}{\p t_0}, 
\eeq
with $\phi_{j,k}(\{\lambda_i\})$ defined with the help of ``elliptic Faber 
functions''. We show that the system (\ref{intro1}) is compatible and the 
associated diagonal metric is of Egorov type. The system (\ref{intro1}) can be solved
by the generalized hodograph method developed by Tsarev in \cite{tsa:90}.
For the general theory of equations of hydrodynamic type see also
\cite{DN,Pavlov,FK}.

The paper is organized as follows. In section 2 we review the algebraic and elliptic formulations 
of the dDKP hierarchy. In section 3 we define the $N$-variable reductions with the help of
a system of elliptic L\"owner equations. Their compatibility condition (the elliptic analogue 
of the Gibbons-Tsarev system) is derived in section 4. Section 5 is devoted to the generalized 
hodograph method. In section 6 we prove that the associated diagonal metric is of Egorov
type and find its potential function. Finally, in section 7 
we discuss conserved quantities. Some long calculations with elliptic and theta functions 
are contained in the appendices.

\section{The dispersionless DKP hierarchy}

We begin with the algebraic form of the dDKP hierarchy.
In what follows we use the differential operator
\beq\label{E5}
\nabla (z)=\p_{t_0}+ D(z)
\eeq
which in the dDKP case is more convenient than $D(z)$. 
Introducing the functions
\beq\label{D4}
p(z)=z-\p_{t_1} \nabla (z)F,
\qquad
w(z)=z^2 e^{-2\p_{t_0}\nabla (z)F},
\eeq
we can rewrite equations (\ref{D1}), (\ref{D2}) in a more compact form
\beq\label{D1a}
e^{D(z)D(\zeta )F}\left (1-\frac{1}{w(z)w(\zeta )}\right )=
\frac{p(z)-p(\zeta )}{z-\zeta}\,,
\eeq
\beq\label{D2a}
e^{-D(z)D(\zeta )F + 2\p_{t_0}^2F}
\, \, \frac{w(z)-w(\zeta )}{z-\zeta}=p(z)+p(\zeta ).
\eeq
Multiplying the two equations, we get the relation
$$
p^2(z)-e^{2F_{00}}\Bigl (w(z)+w^{-1}(z)\Bigr )=
p^2(\zeta )-e^{2F_{00}}\Bigl (w(\zeta )+w^{-1}(\zeta )\Bigr )
$$
from which it follows that $p^2(z)-e^{2F_{00}}\Bigl (w(z)+w^{-1}(z)\Bigr )$
does not depend on $z$ (here and below we use the short-hand notation $F_{mn}=\frac{\p^2 F}{\p t_m \p t_n}$). 
Tending $z$ to infinity, we find that this 
expression is equal to 
$F_{02} -2F_{11} -F_{01}^2$.
Therefore, we conclude that $p(z), w(z)$ satisfy the
algebraic equation \cite{Takasaki09}
\beq\label{D5}
p^2(z)=R^2 \Bigl (w(z)+w^{-1}(z)\Bigr )+V\,,
\eeq
where
\beq\label{D6}
R=e^{F_{00}}, \qquad 
V=F_{02} -2F_{11} -F^2_{01}.
\eeq
are real numbers depending on the times ($R$ is positive).
This equation defines an elliptic curve, with
$w$, $p$ being algebraic functions on this curve. 

A natural further step is to uniformize the curve through  
elliptic functions. This provides the elliptic formulation of the dDKP
hierarchy which was suggested in \cite{akh-zab:14-1}, see also 
\cite{akh-zab:14-2}.
To this end, we use the standard Jacobi
theta functions $\theta_a (u)=\theta_a (u, \tau )$ ($a=1,2,3,4$).
(Their definition is given in Appendix A.)
The elliptic parametrization of (\ref{D5}) is as follows:
\beq\label{E1}
w(z)=\frac{\theta_4^2(u(z))}{\theta_1^2(u(z))}\,, \qquad
p(z)=\gamma \, \theta_4^2(0)\, \frac{\theta_2(u(z))\,
\theta_3(u(z))}{\theta_1(u(z))\, \theta_4(u(z))}\,,
\eeq
where $u(z)=u(z, {\bf t})$ is some function 
of $z$, $\gamma$ is a $z$-independent
factor, and
\beq\label{E2}
R=\gamma\, \theta_2(0)\, \theta_3(0)\,, \qquad
V=-\gamma^2 \Bigl ( \theta_2^4(0)+\theta_3^4(0)\Bigr ).
\eeq
Here $\gamma$ is an arbitrary real parameter but we will see that
it can not be put equal to a fixed number like 1 because it is a 
dynamical variable, as well as the modular parameter $\tau$:
$\gamma =\gamma ({\bf t})$, $\tau =\tau ({\bf t})$.
The reality 
of the coefficients  $R^2$, $V$ 
implies certain restrictions on possible 
values of $\tau$.
The sufficient condition is that $\tau$ is purely imaginary,
which we assume in what follows.
It is convenient to normalize $u(z)$ by the condition 
$u(\infty )=0$, with the expansion around $\infty$ being
\beq\label{E3}
u(z, {\bf t})=\frac{c_1({\bf t})}{z}+\frac{c_2({\bf t})}{z^2}+\ldots 
\eeq
with real coefficients $c_i$.

The
equations (\ref{D1a}), (\ref{D2a}) are then represented as a single
equation:
\beq\label{E4}
\left (z_1^{-1}-z_2^{-1}\right ) e^{\nabla (z_1) \nabla (z_2)F}
=\frac{\theta_1(u(z_1)\! -\! u(z_2))}{\theta_4(u(z_1)\! -\! u(z_2))}\,.
\eeq
The limit $z_2\to \infty$ in (\ref{E4}) 
gives the definition of the 
function $u(z)$:
\beq\label{E6}
e^{\p_{t_0}\nabla (z)F}=z\,
\frac{\theta_1(u(z))}{\theta_4(u(z))}
\eeq
(equivalent to the first formula in (\ref{E1})).
The $z\to \infty$ limit of equation (\ref{E6}) yields:
\beq\label{E8}
e^{F_{00}}=R=\pi c_1 \, \theta_2(0)\theta_3(0),
\eeq
hence
$
c_1({\bf t})=\gamma ({\bf t})/\pi$.

Another useful form of equation (\ref{E4}) can be obtained
by passing to logarithms and applying $\nabla (z_3)$ to the both sides. 
It is convenient to introduce the function
\beq\label{E10}
S(u , \tau ):=\log \frac{\theta_1(u ,\tau )}{\theta_4(u ,\tau )}\,,
\eeq
which has the following quasiperiodicity properties:
$S(u+1,\tau )=S(u,\tau )+i\pi$, 
$S(u+\tau ,\tau )=S(u,\tau )$.
In what follows we write simply $S(u)=S(u, \tau )$.
In terms of this function, the equation (\ref{E4}) means that
$\nabla (z_3) S (u(z_1)-u(z_2)) =\nabla (z_3)\nabla (z_2)\nabla (z_1)F$ is symmetric
under permutations of $z_1, z_2, z_3$:
\beq\label{E11}
\nabla (z_1) S (u(z_2)-u(z_3)) =
\nabla (z_2) S (u(z_1)-u(z_3)) =
\nabla (z_3) S (u(z_1)-u(z_2)) .
\eeq
In particular, as $z_3\to \infty$ we get
\beq\label{E12}
\nabla (z_1) S (u(z_2)) =\p_{t_0}S (u(z_1)-u(z_2)).
\eeq
Thus we may say that the pair $u(z, {\bf t})$, $\tau ({\bf t})$ satisfying this equation
is a solution to the dDKP hierarchy.

In order to connect this with the algebraic formulation, we note that
\beq\label{E13}
\log w(z)=-2S(u(z) ), \qquad
p(z)=c_1 S'(u(z)),
\eeq
where $S'(u)\equiv \p_u S(u)$. See \cite{akh-zab:14-1} for details.

\section{From the elliptic L\"owner equation to the dDKP hierarchy}

In this section we prove that a solution of a system of elliptic L\"owner 
equations gives a solution to the dDKP hierarchy (the $N$-variable diagonal reduction).

Let $u=u(z, \{\lambda_i\})$ be a function of $z$ and real variables
$\{\lambda_i\}=\{\lambda_1, \ldots , \lambda_N\}$. 
Consider the system of elliptic L\"owner (Goluzin-Komatu) equations
\beq\label{f1}
\frac{\p u}{\p \lambda_j}=\frac{1}{4\pi {\rm i}}\Bigl (
-\zeta_1 (u+\xi_j)-\zeta_4 (u+\xi_j)+\zeta_1(\xi_j)+\zeta_4(\xi_j)\Bigr )
\frac{\p \tau}{\p \lambda_j},
\eeq
where $\zeta_a (x)= \zeta_a(x, \tau) =\p_x\log \theta_a(x, \tau)$ are analogues of the 
Weierstrass' zeta function, $\xi_j$ and $\tau$ are functions of $\{\lambda_i\}$:
$\xi_j=\xi_j(\{\lambda_i\})$, $\tau =\tau (\{\lambda_i\})$. We assume that 
$\xi_j$ are real-valued functions.

Let us find $\p_{\lambda_j}S(u(z_1)-u(z_2))$:
$$
\p_{\lambda_j}S(u_1-u_2)=
S'(u_1-u_2)\Bigl (\frac{\p u_1}{\p \lambda_j}-\frac{\p u_2}{\p \lambda_j}\Bigr )
+\dot S(u_1-u_2)\frac{\p \tau}{\p \lambda_j}\,,
$$
where we abbreviate $u_i \equiv u(z_i)$ and $\dot S(u)=\p_{\tau}S(u, \tau )$.
Plugging here the elliptic L\"owner equations (\ref{f1}) and the formula 
\beq\label{f2}
2\pi {\rm i}\dot S(u)=S'(u)\zeta_2(u, \tau)+\frac{\pi^2}{2}\, \theta_4^4(0, \tau)
\eeq
(see \cite{akh-zab:14-1} for the proof and \cite{Takasaki01,ZZ12} for the proofs 
of similar formulae), we have:
\beq\label{f201}
\begin{array}{c}
\p_{\lambda_j}S(u_1\! -\! u_2)=
\displaystyle{\frac{1}{4\pi {\rm i}}S'(u_1-u_2) \Bigl [
-\zeta_1 (u_1+\xi_j)-\zeta_4 (u_1+\xi_j)+\zeta_1 (u_2+\xi_j)+\zeta_4 (u_2+\xi_j)}
\\  \\
 \displaystyle{ +\, 2\zeta_2(u_1-u_2)+\frac{\pi^2 \theta_4^4(0, \tau)}{S'(u_1-u_2)}
\Bigr ] \frac{\p \tau}{\p \lambda_j}\phantom{aaaaaaaaaa}}
\\ \\
=\displaystyle{ \frac{1}{4\pi {\rm i}} S'(u_1 +\xi_j)S'(u_2+\xi_j)
\frac{\p \tau}{\p \lambda_j}}  ,
\end{array}
\eeq
where we have used the identity (A15) from \cite{akh-zab:14-1}.
In particular, tending $z_2\to \infty$, we get
\beq\label{f202}
\p_{\lambda_j}S(u(z))=\frac{1}{4\pi {\rm i}} S'(\xi_j)S'(u(z)+\xi_j)
\frac{\p \tau}{\p \lambda_j}.
\eeq

Using the above functions $u(z, \{\lambda_i\})$, $\tau (\{\lambda_i\})$ let us construct
a solution $u(z)$, $\tau$ to the dDKP hierarchy which depends on times through the 
$\lambda_i$'s: $u(z, {\bf t})=u(z, \{\lambda_i ({\bf t})\})$, $\tau ({\bf t})=\tau (\{\lambda_i({\bf t})\})$.
This is called $N$-variable reduction of the hierarchy.
In the case of the reduction equation (\ref{E12}) reads
$$
\sum_{j=1}^N\nabla (z_1)\lambda_j \cdot
\p_{\lambda_j}S(u(z_2))=\sum_{j=1}^N \p_{t_0}\lambda_j \cdot
\p_{\lambda_j}S(u(z_1)-u(z_2))
$$
Plugging here (\ref{f201}), (\ref{f202}), we have:
\beq\label{f203}
\sum_{j=1}^N\nabla (z_1)\lambda_j \cdot S'(\xi_j)S'(u(z_2)\! +\! \xi_j)
\frac{\p \tau}{\p \lambda_j}
=\sum_{j=1}^N\p_{t_0}\lambda_j \cdot S'(u(z_1)\! +\! \xi_j)S'(u(z_2)\! +\! \xi_j)
\frac{\p \tau}{\p \lambda_j}.
\eeq
Now we see that if we introduce the dependence of the $\lambda_j$'s on ${\bf t}$ 
by means of the equation
\beq\label{f3}
\nabla (z)\lambda_j=\frac{S'(u(z)+\xi_j)}{S'(\xi_j)}\, \frac{\p \lambda_j}{\p t_0},
\eeq
eq. (\ref{f203}) is satisfied identically. 
This means that the functions $u(z,{\bf t})=u(z, \{\lambda_i(\bf t)\})$,
$\tau ({\bf t})=\tau (\{\lambda_i(\bf t)\})$ obey the dDKP hierarchy.

Equation (\ref{f3}) contains an infinite system of partial differential equations 
of hydrodynamic type. To write them out explicitly, we introduce elliptic Faber functions 
$\Phi_k (u)$ via the expansion 
$\displaystyle{S(u(z)+w)=S(w)+\sum_{k=1}^{\infty}\frac{z^{-k}}{k}\, \Phi _k(w)}$ or
\beq\label{f4}
S'(u(z)+w)=S'(w)+\sum_{k=1}^{\infty}\frac{z^{-k}}{k}\, \Phi '_k(w)
\eeq
(here $\Phi '_k(w)=\p_w\Phi _k(w)$). Then the system (\ref{f3}) reads
\beq\label{f5}
\frac{\p \lambda _j}{\p t_k}=\phi_{j,k}(\{\lambda_i\})\, \frac{\p \lambda _j}{\p t_0}\,, \qquad
\phi_{j,k}=\frac{\Phi '_k(\xi_j)}{S'(\xi_j)},
\eeq
which is an infinite diagonal system of partial differential equations of hydrodynamic type.
The $\lambda_j$'s play the role of the Riemann invariants. 
Note that $\Phi '_k(w)$ depends on the $\lambda_j$'s through $\tau$ and $u(z)$.
The generating function of $\phi_{j,k}(\{\lambda_i\})$ is obtained from (\ref{f4}):
\beq\label{f6}
Q\Bigl (u(z, \{\lambda_i\}, \xi_j(\{\lambda_i\}), \tau (\{\lambda_i\})\Bigr )=
1+\sum_{k\geq 1}\phi_{j,k}(\{\lambda_i\})\frac{z^{-k}}{k}, \qquad
Q(u, \xi , \tau )=\frac{S'(u+\xi , \tau )}{S'(\xi , \tau )}.
\eeq
It is convenient to put $\phi_{i,0}=1$.

\section{The Gibbons-Tsarev system}

The Gibbons-Tsarev system is the compatibility condition for the system 
of elliptic L\"owner equations (\ref{f1}):
\beq
\label{gt2}
\begin{array}{lll}
\displaystyle{\frac{\p u}{\p \lambda_j}}&=&\displaystyle{\frac{1}{4\pi {\rm i}}\Bigl (
-\zeta_1 (u+\xi_j, \tau)-\zeta_4 (u+\xi_j,\tau )+\zeta_1(\xi_j,\tau )+\zeta_4(\xi_j, \tau )\Bigr )
\frac{\p \tau}{\p \lambda_j}}
\\ &&\\
&=&\displaystyle{\frac{1}{4\pi {\rm i}}\Bigl (
-\zeta_1 (u+\xi_j, \tau ')+\zeta_1(\xi_j,\tau ' )\Bigr )
\frac{\p \tau}{\p \lambda_j}}.
\end{array}
\eeq
Here and below we abbreviate $\tau ' =\frac{\tau}{2}$.  The
compatibility condition is
$$
F_{jk}(u):= \frac{\p }{\p \lambda_j}\, \frac{\p u}{\p \lambda_k}-
\frac{\p }{\p \lambda_k}\, \frac{\p u}{\p \lambda_j}=0.
$$
The left hand side is of the form 
\beq\label{gt31}
F_{jk}(u)=F_{jk}^{(1)}\frac{\p \xi_k}{\p \lambda_j}\, \frac{\p \tau }{\p \lambda_k}
-F_{kj}^{(1)}\frac{\p \xi_j}{\p \lambda_k}\, \frac{\p \tau }{\p \lambda_j}
+F_{jk}^{(2)}\frac{\p^2 \tau}{\p \lambda_j \p \lambda_k}
+G_{jk}\frac{\p \tau}{\p \lambda_j}\, \frac{\p \tau}{\p \lambda_k}.
\eeq
The coefficients are:
$$
F_{jk}^{(1)}=\frac{1}{4\pi {\rm i}}\Bigl ( \wp_1 (u+\xi_k), \tau ')-\wp_1 (\xi_k, \tau ')\Bigr ),
$$
$$
F_{jk}^{(2)}=\frac{1}{4\pi {\rm i}}\Bigl (-\zeta_1 (u+\xi_k, \tau ')+
\zeta_1(\xi_k, \tau ')+\zeta_1 (u+\xi_j, \tau ')-
\zeta_1(\xi_j, \tau ')\Bigr ),
$$
$$
G_{jk}=\frac{1}{2(4\pi {\rm i})^2}\Bigl (
\wp_1'(u+\xi_k, \tau ')-\wp_1'(u+\xi_j, \tau ')-\wp_1'(\xi_k, \tau ')+\wp_1'(\xi_j, \tau ')\Bigr )
$$
$$
+\,\, \frac{1}{(4\pi {\rm i})^2}\Bigl (\zeta_1 (u+\xi_k, \tau')-\zeta_1 (u+\xi_j, \tau')+
\zeta_1(\xi_j, \tau ')\Bigr ) \wp_1 (u+\xi_k, \tau ')
$$
$$
-\,\, \frac{1}{(4\pi {\rm i})^2}\Bigl (\zeta_1 (u+\xi_j, \tau')-\zeta_1 (u+\xi_k, \tau')+
\zeta_1(\xi_k, \tau ')\Bigr ) \wp_1 (u+\xi_j, \tau ')
$$
$$
+\,\, \frac{1}{(4\pi {\rm i})^2}\Bigl (-\zeta_1(\xi_k, \tau ')\wp_1(\xi_k, \tau ')+
\zeta_1(\xi_j, \tau ')\wp_1(\xi_j, \tau ')\Bigr ),
$$
where $\wp_a (x, \tau )=-\p_x \zeta _a(x, \tau )$, 
$\wp_a' (x, \tau )=\p_x \wp _a(x, \tau )$. 
Here we have used the equation
\beq\label{gt32}
4\pi {\rm i}\, \frac{\p \zeta_1(u, \tau ')}{\p \tau}=
-\zeta_1(u, \tau ')\wp_1(u, \tau ')-\frac{1}{2}\, \wp_1'(u, \tau')
\eeq
(equivalent to eq. (A11) in \cite{akh-zab:14-1}).
Note that
\begin{itemize}
 \item $\wp_{1}(u,\tau')$ is an elliptic function with periods $1$ and
       $\tau'$. Hence $\wp '_{1}(u, \tau ')$ is also an elliptic
       function with the same periods.

 \item $\zeta_{1}(u+1,\tau')=\zeta_{1}(u,\tau')$ and
       $\zeta_{1}(u+\tau',\tau')=\zeta_{1}(u,\tau')-2\pi
       {\rm i}$, which implies that
       $\zeta_{1}(u+\xi_k,\tau')-\zeta_{1}(u+\xi_j,\tau')$ is an
       elliptic function of $u$ with periods $1$ and $\tau'$.
\end{itemize}
Therefore, $F_{jk}^{(1)}$, $F_{jk}^{(2)}$, $G_{jk}$ and consequently
$F_{jk}(u)$ are all elliptic functions of $u$ with periods $1$ and
$\tau'$. Possible poles in the parallelogram spanned by $1$ and
$\tau'$ are at $u=-\xi_k$ and $u=-\xi_j$.

The formulae of expansions around $u=0$ are:
\beq\label{gt4}
\zeta_1(u, \tau ')=\frac{1}{u}+O(u), \qquad
\wp_1(u, \tau ')=\frac{1}{u^2}+O(1), \qquad
\wp '_1(u, \tau ')=-\frac{2}{u^3}+O(u).
\eeq
Using them, we have the expansions near $u=-\xi_k$:
\beq\label{Fjk1}
    F_{jk}^{(1)} = 
    \frac{1}{4\pi {\rm i}}\left (
     \frac{1}{(u+\xi_k)^2} + O(1)\right ),
\eeq
\beq\label{Fjk2}
    F_{jk}^{(2)} = 
    \frac{1}{4\pi {\rm i}}\left (
     -\frac{1}{u+\xi_k} + O(1)\right ),
\eeq
\beq\label{Gjk}
    G_{jk} =\frac{1}{(4\pi {\rm i})^2}\left (\frac{\zeta_1(\xi_j, \tau ')
    -\zeta_1 (-\xi_k+\xi_j, \tau ')}{(u+\xi_k)^2}\, +\frac{2\wp_1(\xi_j-\xi_k, \tau ')}{u+\xi_k}+
    O(1)
    \right ).
\eeq
Substituting them into (\ref{gt31}), we can expand $F_{jk}(u)$ around $u=-\xi_k$ as
$$
F_{jk}(u)=\frac{f_2}{(u+\xi_k)^2}+\frac{f_1}{u+\xi_k}+O(1),
$$
where
$$
f_2=\frac{1}{4\pi {\rm i}}\, \frac{\p \tau}{\p \lambda_k}\left (
\frac{\p \xi_k}{\p \lambda_j}+\frac{1}{4\pi {\rm i}}
\Bigl (\zeta_1(\xi_j, \tau ')
    -\zeta_1 (-\xi_k+\xi_j, \tau ')\Bigr )
\frac{\p \tau}{\p \lambda_j}\right ),
$$
$$
f_1=\frac{1}{4\pi {\rm i}}\left (\frac{2}{4\pi {\rm i}}
\frac{\p \tau}{\p \lambda_j}\frac{\p \tau}{\p \lambda_k}\, \wp_1(\xi_j-\xi_k, \tau ')-
\frac{\p^2 \tau }{\p \lambda_j \p \lambda_k}\right ).
$$
Therefore, if 
\beq\label{GT1}
\frac{\p \xi_k}{\p \lambda_j}=\frac{1}{4\pi {\rm i}}\, \Bigl (
\zeta_1(-\xi_k+\xi_j, \tau ')-\zeta_1(\xi_j, \tau ')\Bigr )\, \frac{\p \tau}{\p \lambda_j},
\eeq
\beq\label{GT2}
\frac{\p^2 \tau}{\p \lambda_k \p \lambda_j}=\frac{1}{2\pi {\rm i}}\,
\wp _1(\xi_k-\xi_j, \tau ')\frac{\p \tau}{\p \lambda_k}\frac{\p \tau}{\p \lambda_j}
\eeq
for all $j=1, \ldots , N$, $j\neq k$, then $F_{jk}(u)$ is regular at $u=-\xi_k$. Similarly,
if these equations with $j$ and $k$ exchanged are satisfied for all
$k=1,\ldots,N$, $k\neq j$, then $F_{jk}(u)$ is regular at $u=-\xi_j$. 

Assume that equations (\ref{GT1}) and (\ref{GT2}) hold for all
$j,k=1,\ldots ,N$, $j\neq k$. Then $F_{jk}(u)$ is a regular elliptic function, which is
nothing but a constant. It is easy to see that $F_{jk}(0)=0$. So under the
conditions (\ref{GT1}) and (\ref{GT2}), $F_{jk}(u)=0$ which means that the elliptic
L\"owner system (\ref{f1}) is compatible.

The system of 
equations (\ref{GT1}), (\ref{GT2}) is the elliptic analogue of the famous Gibbons-Tsarev system
\cite{GT1,GT2}. They already appeared in the literature 
\cite{ode-sok:09,ode-sok:10,ode-sok:09-1}.

\section{Generalized hodograph method}

In the previous sections we have reduced the dDKP hierarchy to
the system of elliptic L\"owner equations and the auxiliary equations
\begin{equation}\label{g0}
    \frac{\p \lambda_i ({\bf t})}{\p t_n}
    =
    \phi_{i,n}(\{\lambda_j({\bf t})) \frac{\p\lambda_i({\bf t})}{\p t_0},
\end{equation}
where $\phi_{i,n}$ are as in (\ref{f5}).
We are going to show that 
this system of the first order partial differential equations is consistent and can be solved by
Tsarev's generalized hodograph method \cite{tsa:90}.

As is easy to see, 
the compatibility condition of the system (\ref{g0}) is 
$$
\frac{\p_{\lambda_j}\phi_{i,n}}{\phi_{j,n}-\phi_{i,n}}=
\frac{\p_{\lambda_j}\phi_{i,n'}}{\phi_{j,n'}-\phi_{i,n'}}\qquad \mbox{for all $i\neq j$, $n, n'$}.
$$
In other words, we should 
show that 
\beq\label{g1}
\Gamma_{ij}:=\frac{\p_{\lambda_j}\phi_{i,n}}{\phi_{j,n}-\phi_{i,n}}
\eeq
does not depend on $n$, i.e., (\ref{g1}) holds for all $n$ 
simultaneously. This is equivalent to the statement that the ratio
\beq\label{g2}
\frac{\p_{\lambda_j}Q(u(z), \xi_i, \tau )}{Q(u(z), \xi_j, \tau )-Q(u(z), \xi_i, \tau )},
\eeq
where $Q$ is the generating function (\ref{f6}), is independent of $z$.
The independence of (\ref{g2}) of $z$ is proved in Appendix B by a direct calculation, 
and the coefficients $\Gamma_{ij}$ are found. The result is
\beq\label{g3}
\Gamma_{ij}=-\, \frac{1}{4\pi {\rm i}}\, \frac{S'(\xi_j)}{S'(\xi_i)}
\, S''(\xi_i-\xi_j)\, \frac{\p \tau}{\p \lambda_j}.
\eeq

\begin{proposition}
Consider the following system for $R_i=R_i(\{\lambda _j\})$, $i=1, \ldots , N$:
\beq\label{g4}
\frac{\p R_i}{\p \lambda_j}=\Gamma_{ij}(R_j-R_i), \qquad
i,j=1, \ldots , N, \quad i\neq j,
\eeq
where $\Gamma_{ij}$ is defined as in (\ref{g3}) (when $N=1$, the condition (\ref{g4}) is void).
Then the following holds.
\begin{itemize}\item[(i)] The system (\ref{g4}) is compatible in the sense of \cite{tsa:90}.
\item[(ii)]
Assume that $R_i$ satisfy the system (\ref{g4}). If $\lambda_i({\bf t})$ is defined implicitly by the
hodograph relation
\beq\label{g5}
t_0+\sum_{n\geq 1}\phi_{i,n}(\{\lambda_j\})t_n = R_i(\{\lambda_j\}),
\eeq
then $\lambda_j({\bf t})$ satisfy (\ref{g0}).
\end{itemize}
\end{proposition}
For the proof of statement (i) of the proposition we note that $\Gamma_{ij}$ (\ref{g3})
can be expressed as logarithmic derivative of a function as follows:
\beq\label{c2}
\Gamma_{ij}=\frac{1}{2}\, \frac{\p }{\p \lambda_j}\log g_i,
\eeq
where
\beq\label{c3}
g_i=\frac{1}{4\pi {\rm i}}\, (S'(\xi_i))^2 \, \frac{\p \tau}{\p \lambda_i}.
\eeq
The proof of (\ref{c2}) is given in 
Appendix C.
It then follows that 
\beq\label{c2a}
\frac{\p \Gamma_{ij}}{\p \lambda_k}=\frac{\p \Gamma_{ik}}{\p \lambda_j}, \qquad i\neq j\neq k,
\eeq
which is the Tsarev compatibility condition. This means that the system (\ref{g0}) is semi-Hamiltonian.
The main geometric object associated with a semi-Hamiltonian system is a diagonal metric.
The quantities $g_i=g_{ii}$ are components of this metric while $\Gamma_{ij}=\Gamma_{ij}^{i}$ 
are the corresponding Christoffel symbols.

In fact the compatibility conditions of the system (\ref{g4}) are (\ref{c2a}) 
together with
\beq\label{c101}
\frac{\p \Gamma_{ij}}{\p \lambda_k}=
\Gamma_{ij}\Gamma_{jk}+\Gamma_{ik}\Gamma_{kj}-\Gamma_{ik}\Gamma_{ij},
\qquad i\neq j\neq k.
\eeq
As one can see, (\ref{c2a}) follows from (\ref{c101}) because the right hand side
of it is explicitly symmetric under the permutation of $j$ and $k$.
As is shown in \cite{tsa:90}, (\ref{c101}), in its turn, follows from the definition 
(\ref{g1}) and the condition (\ref{c2a}). In Appendix D we give an independent direct proof of 
(\ref{c101}) starting from the explicit form of the $\Gamma_{ij}$.

The proof of statement (ii) of the proposition is almost 
the same as that of Theorem 10 of Tsarev's paper \cite{tsa:90}. 
The difference from \cite{tsa:90} is that 
the number of independent variables is infinite in our case.
In spite of this difference, Tsarev's method does work.
For completeness, we give the proof here.
(The following argument is the same as in \cite{T13}.)
By differentiating the relation (\ref{g5}) by $t_0$ and $t_k$ we
obtain
\begin{equation}
    \sum_{j=1}^N M_{ij} \frac{\p \lambda_j}{\p t_0} = 1,
    \qquad
    \sum_{j=1}^N M_{ij} \frac{\p \lambda_j}{\p t_k} 
    = \phi_{i,k},
\label{d(hodograph)}
\end{equation}
where
\begin{equation}
    M_{ij} :=
    \frac{\p R_i}{\p \lambda_j}
    -\sum_{n\geq 1}
    \frac{\p \phi_{i,n}}{\p \lambda_j} \, t_n.
\label{def:Mij}
\end{equation}
Because of (\ref{g1}), (\ref{g4}) and the hodograph relation (\ref{g5})
 the above expression becomes
$$
 \begin{array}{ll}
    M_{ij}
    &=\displaystyle{
    \Gamma_{ij}(R_j-R_i)
    -
    \sum_{n\geq 1}
    \Gamma_{ij} (\phi_{j,n}-\phi_{i,n}) t_n}
\\ & \\
    &=\displaystyle{
    \Gamma_{ij}\Bigl(
     (R_j - \sum_{n\geq 1} \phi_{j,n} t_n)
     -
     (R_i - \sum_{n\geq 1} \phi_{i,n} t_n)
    \Bigr) = 0,}
 \end{array}
$$
 if $i\neq j$. (The fact that the coefficients $\Gamma_{ij}$ defined by (\ref{g1}) do not
 depend on $n$ is essential here.) Therefore, (\ref{d(hodograph)})
 reduces to
$$
    M_{ii} \frac{\p \lambda_i}{\p t_0} = 1,
    \qquad
    M_{ii} \frac{\p \lambda_i}{\p t_k} = \phi_{i,k}.
$$
 This proves (\ref{g0}).
 
\section{The metric coefficients $g_i$}

\begin{proposition}
The metric $g_i$ is of Egorov type, i.e., it holds
\beq\label{q1}
\frac{\p g_i}{\p \lambda_k}=\frac{\p g_k}{\p \lambda_i}.
\eeq
\end{proposition}
The proof is very simple:
$$
\frac{\p g_k}{\p \lambda_i}=g_k\frac{\p \log g_k}{\p \lambda_i}
=2g_k\Gamma_{ki}=
-\frac{2}{(4\pi {\rm i})^2}S'(\xi_i)S'(\xi_k)S''(\xi_i-\xi_k)
\frac{\p \tau}{\p \lambda_i}\, \frac{\p \tau}{\p \lambda_k}
$$
which is explicitly symmetric under the permutation of $i$ and $k$.
Here we use (\ref{g3}) and (\ref{c3}).

The relations (\ref{q1}) imply that the quantities $g_i$ have a potential function $G$
such that 
\beq\label{p1}
g_i=\frac{\p G}{\p \lambda _i}.
\eeq
Let us show that
\beq\label{p2}
G=\log R = \frac{\p^2F}{\p t_0^2},
\eeq
where $F$ is the tau-function (free energy) of the dDKP hierarchy (see (\ref{E8})).
The starting point is the system of the elliptic L\"owner equations
\beq\label{p3}
4\pi {\rm i}\, 
 \p_{\lambda_j}u(z)=\Bigl (-\zeta_1(u(z)+\xi_j, \tau')+\zeta_1(\xi_j, \tau ')\Bigr )\frac{\p \tau}{\p \lambda_j},
\eeq
where $u(z)$ has the expansion 
$\displaystyle{
u(z)=\frac{c_1}{z}+\frac{c_2}{z^2}+\ldots}$ as $z\to \infty$.
 Expanding both sides of (\ref{p3}) as $z\to \infty$ and equating the coefficients in front of $z^{-1}$, 
we have
\beq\label{p4}
4\pi {\rm i}\, 
 \p_{\lambda_j}\log c_1 =\wp_1(\xi_j, \tau ')\frac{\p \tau}{\p \lambda_j}.
 \eeq
As is shown in \cite{akh-zab:14-1} (see also equation (\ref{E8}) in the present paper),
$
\log (\pi c_1)=\log R -\log \frac{\theta_2^2(0, \tau ')}{2}
$.
Therefore,
$$
\begin{array}{lll}
\p_{\lambda_j}\log c_1 &=&\p_{\lambda_j}\log R -2\p_{\lambda_j}\log \theta_2(0, \tau ')
\\ && \\
&=&\displaystyle{\p_{\lambda_j}\log R -2\p_{\tau '}\log \theta_2(0, \tau ')\frac{\p \tau '}{\p \lambda_j}}
\\ && \\
&=&\displaystyle{\p_{\lambda_j}\log R -\p_{\tau '}\log \theta_2(0, \tau ')\frac{\p \tau }{\p \lambda_j}}.
\end{array}
$$
From the heat equation $4\pi {\rm i}\,\p_{\tau}\theta_a(x,\tau )=\theta_{a}''(x, \tau)$ it follows that
$$
4\pi {\rm i}\,\p_{\tau '}\log \theta_2(0, \tau ')=-\wp_2(0, \tau ').
$$
Plugging this into (\ref{p4}), we have
$$
4\pi {\rm i}\, \p_{\lambda_j}\log R=\Bigl (\wp_1(\xi_j, \tau ')-\wp_2(0, \tau ')\Bigr )
\frac{\p \tau }{\p \lambda_j}=
(S'(\xi_j))^2 \, \frac{\p \tau }{\p \lambda_j}=4\pi {\rm i} \, g_i
$$
(see (\ref{n12})).

\section{Conserved quantities}

As is shown in \cite{tsa:90}, densities $P$ of the conserved quantities 
$I=\int P dt_0$ for any semi-hamiltonian system $\p_{t_k}{\lambda_j}=
\phi_{j,k}\p_{t_0}{\lambda_j}$ satisfy the linear differential equation
\beq\label{cq1}
\frac{\p^2 P}{\p \lambda_i \p \lambda_j}=\Gamma_{ij}
\frac{\p P}{\p \lambda_i}+\Gamma_{ji}
\frac{\p P}{\p \lambda_j} \qquad (i\neq j)
\eeq
and any solution to this equation gives a conserved quantity. Indeed, substituting (\ref{g1})
for $\Gamma_{ij}$, we have
$$
\p_{\lambda_i}\p_{\lambda_j}P=\frac{\p_{\lambda_j}
\phi_{i,n}\p_{\lambda_i}P}{\phi_{j,n}-\phi_{i,n}}+
\frac{\p_{\lambda_i}
\phi_{j,n}\p_{\lambda_j}P}{\phi_{i,n}-\phi_{j,n}}
$$
which is equivalent to
\beq\label{cq2}
\p_{\lambda_j}\Bigl ( \p_{\lambda_i}P \, \phi_{i,n}\Bigr )=
\p_{\lambda_i}\Bigl ( \p_{\lambda_j}P \, \phi_{j,n}\Bigr ).
\eeq
This means that there exists a function $A_n$ such that
$\p_{\lambda_i}P \, \phi_{i,n}=\p_{\lambda_i}A_n$. Then
$$
\frac{\p P}{\p t_n}=\sum_{i=1}^N \p_{\lambda_i}P \, \p_{t_n}\lambda_i =
\sum_{i=1}^N \p_{\lambda_i}P \, \phi_{i,n}\p_{t_0}\lambda_i =
\sum_{i=1}^N \p_{\lambda_i}A_n \, \p_{t_0}\lambda_i 
=\frac{\p A_n}{\p t_0}
$$
which means that $P$ is indeed the density of a conserved quantity.

In Appendix E we prove that the function $S(u(z))$ satisfies equation (\ref{cq1}):
\beq\label{cq3}
\frac{\p^2 S(u(z))}{\p \lambda_i \p \lambda_j}=\Gamma_{ij}
\frac{\p S(u(z))}{\p \lambda_i}+\Gamma_{ji}
\frac{\p S(u(z))}{\p \lambda_j}
\eeq
provided $u(z)$ obeys the elliptic L\"owner equations.  
Therefore, $S(u(z))$ is the generating function for densities of conserved quantities.
According to the definition (\ref{E6}) of the function $u(z)$,
\beq\label{cq4}
S(u(z))=-\log z +F_{00}+\sum_{n\geq 1} \frac{z^{-n}}{n}\, F_{0n},
\eeq
so the densities are $F_{0n}$, $n\geq 0$. 
(The fact that densities of conserved quantities are expressed through second order
logarithmic derivatives of tau-function is common for integrable hierarchies,
see, e.g., \cite{Sato-Sato,Takebe90}.)
According to (\ref{f202}) we have
$$
\p_{\lambda_j}S(u(z))=g_j\, \frac{S'(u(z)+\xi_j)}{S'(\xi_j)}=
g_j\left ( 1+\sum_{n\geq 1}\phi_{j, n}\frac{z^{-n}}{n}\right ).
$$
Comparing with the $\lambda_j$-derivative of (\ref{cq4}), we conclude that
\beq\label{cq5}
g_j\phi_{j,n}=\p_{\lambda_j}F_{0n}, \qquad n\geq 0,
\eeq
which generalizes (\ref{p1}), (\ref{p2}) (obtained at $n=0$).

\section{Conclusion}
We have found sufficient conditions for $N$-variable diagonal reduction of the dDKP hierarchy
(the dispersionless limit of the Pfaff lattice) in the elliptic 
parametrization. The reduction is given by $N$ elliptic
L\"owner equations (\ref{f1}) for a function $u(z, \lambda_1, \ldots , \lambda_N)$
supplemented by a diagonal system of hydrodynamic type (\ref{f5}) 
for the variables 
$\lambda_j$, $j=1, \ldots , N$. We have derived compatibility conditions for the 
elliptic
L\"owner equations which are elliptic analogues of the Gibbons-Tsarev equations and
have
proved solvability of the hydrodynamic type system by means of the 
generalized hodograph method. The associated diagonal metric is proved to be of the
Egorov type.

\section*{Appendices}

\subsection*{Appendix A: necessary functions and identities}
\def\theequation{A\arabic{equation}}
\setcounter{equation}{0}

The Jacobi's theta-functions $\theta_a (u)=
\theta_a (u,\tau )$, $a=1,2,3,4$, are defined by the formulas
\beq\label{Bp1}
\begin{array}{l}
\theta _1(u)=-\displaystyle{\sum _{k\in \z}}
\exp \left (
\pi {\rm i} \tau (k+\frac{1}{2})^2 +2\pi {\rm i}
(u+\frac{1}{2})(k+\frac{1}{2})\right ),
\\
\theta _2(u)=\displaystyle{\sum _{k\in \z}}
\exp \left (
\pi {\rm i} \tau (k+\frac{1}{2})^2 +2\pi {\rm i}
u(k+\frac{1}{2})\right ),
\\
\theta _3(u)=\displaystyle{\sum _{k\in \z}}
\exp \left (
\pi {\rm i} \tau k^2 +2\pi {\rm i} u k \right ),
\\
\theta _4(u)=\displaystyle{\sum _{k\in \z}}
\exp \left (
\pi {\rm i} \tau k^2 +2\pi {\rm i}
(u+\frac{1}{2})k\right ),
\end{array}
\eeq where $\tau$ is a complex parameter (the modular parameter) 
such that ${\rm Im}\, \tau >0$. The function 
$\theta_1(u)$ is odd, the other three functions are even.
The infinite product representation for the $\theta_1(u)$ reads: 
\beq
\label{infprod} \theta_1(u)={\rm i}\,\mbox{exp}\, 
\Bigl ( \frac{{\rm i}\pi \tau}{4}-{\rm i}\pi u\Bigr ) \prod_{k=1}^{\infty} 
\Bigl (
1-e^{2\pi {\rm i} k\tau }\Bigr ) \Bigl ( 1-e^{2\pi {\rm i} ((k-1)\tau +u)}\Bigr ) 
\Bigl ( 1-e^{2\pi {\rm i} (k\tau -u)}\Bigr ). 
\eeq 
We also mention the identity
\beq\label{theta1prime}
\theta_1'(0)=\pi \theta_2(0) \theta_3(0) \theta_4(0).
\eeq
Many useful identities for the theta functions can be found in \cite{KZ1}.

All formulas for derivatives of elliptic functions with respect 
to the modular parameter follow from the ``heat equation'' satisfied
by the theta-functions:
\beq\label{heat}
4\pi {\rm i} \, \p_{\tau}\theta_a(u)=\p_u^2 \theta_a(u).
\eeq

In the main text we use the functions
$$
\zeta_a (x, \tau)=\frac{\p }{\p x}\log \theta_a(x, \tau), 
\qquad
\wp_a(x, \tau )=-\frac{\p }{\p x}\zeta_a (x, \tau), \qquad a=1,2,3,4.
$$
Obviously, $\zeta_a$ are odd functions. In particular, 
$\zeta_1(x, \tau)=\frac{1}{x}+O(x)$ as $x\to 0$ and $\zeta_a(0, \tau)=0$ for $a=2,3,4$.

Let us introduce the function
\beq\label{n1}
S(x)=\log \frac{\theta_1(x, \tau)}{\theta_4(x,\tau )}.
\eeq
We denote $\p_x S(x)=S'(x)$, $\p^2_x S(x)=S''(x)$,
$\p_{\tau}S(x)=\dot S(x)$. One can prove the following formulae (here and below 
$\tau' \equiv \frac{\tau}{2}$):
\beq\label{n2}
\begin{array}{lll}
S'(x)&=&\displaystyle{\pi \theta_4^2(0, \tau )\, 
\frac{\theta_2(x, \tau)\theta_3(x,\tau )}{\theta_1(x, \tau)\theta_4(x,\tau )}}
\\ &&\\
&=&\displaystyle{\pi \theta_3(0,\tau')\theta_4(0,\tau')\, \frac{\theta_2(x, \tau')}{\theta_1(x, \tau')},}
\end{array}
\eeq
\beq\label{n3}
\begin{array}{lll}
S''(x)&=&\displaystyle{-\pi^2\theta_2^2(0, \tau)\theta_3^2(0, \tau)\theta_4^3(0, \tau)\,
\frac{\theta_4(2x, \tau )}{\theta_1^2(x, \tau)\theta_4^2(x, \tau)}}
\\ &&\\
&=&\displaystyle{-\pi^2 \theta_3(0,\tau')\theta_4(0,\tau')\theta_2^2(0,\tau')\,
\frac{\theta_3(x, \tau')\theta_4(x, \tau')}{\theta_1^2(x,\tau')},}
\end{array}
\eeq
\beq\label{n4}
2\pi {\rm i}\dot S(x)=S'(x)\zeta_2 (x, \tau)+\frac{\pi^2}{2}\theta_4^4(0, \tau),
\eeq
\beq\label{n5}
2\pi {\rm i}\dot S'(x)=S''(x)\zeta_2(x, \tau)-S'(x)\wp_2(x, \tau ).
\eeq
It is clear from (\ref{n2}), (\ref{n3}) that
$
S'(x+1)=S'(x)$, $S'(x+\tau ')=-S'(x)$,
$S''(x+1)=S''(x)$, $S''(x+\tau ')=-S''(x)
$.
Note that
\beq\label{n6a}
\begin{array}{c}
S'(x)S'(x+\frac{1}{2})=-\pi^2\theta_4^4(0, \tau).
\end{array}
\eeq
As $x\to 0$, we have:
$$
S'(x)=\frac{1}{x}+O(x), \qquad S''(x)=-\frac{1}{x^2}+O(1).
$$
We mention the identities
\beq\label{n6}
S'(x)S''(x)=\frac{1}{2}\, \wp '_1(x, \tau '),
\eeq
\beq\label{n8}
S'(x)=2\zeta_1(x, \tau)-\zeta_1(x, \tau ').
\eeq
It immediately follows from here that
\beq\label{n9}
\begin{array}{c}
2\zeta_2(x, \tau)-\zeta_2(x, \tau ')=S'(x+\frac{1}{2}),
\end{array}
\eeq
\beq\label{n10}
\begin{array}{c}
2\wp _2(x, \tau)-\wp _2(x, \tau ')=-S''(x+\frac{1}{2}).
\end{array}
\eeq
We also need the standard identity
\beq\label{n11}
\wp_2(x,\tau)-\wp_2(y, \tau)=\frac{(\theta_1'(0,\tau))^2 
\theta_1(x-y,\tau)\theta_1(x+y,\tau)}{\theta_2^2(x,\tau)\theta_2^2(y,\tau)}.
\eeq
A particular case is
\beq\label{n12}
\wp_1(x, \tau ')-\wp_2(0, \tau ')=(S'(x))^2.
\eeq

\subsection*{Appendix B: the coefficients $\Gamma_{ij}$}
\def\theequation{B\arabic{equation}}
\setcounter{equation}{0}

Here we show that 
\beq\label{g1b}
\Gamma_{ij}:=\frac{\p_{\lambda_j}\phi_{i,n}}{\phi_{j,n}-\phi_{i,n}}
\eeq
does not depend on $n$, i.e., holds for all $n$ 
simultaneously, and find the coefficients 
$\Gamma_{ij}$. Passing to the generating function of $\phi_{i,n}$,
$$
Q(u(z), \xi_i)=1+\sum_{n\geq 1}\phi_{i,n}\frac{z^{-n}}{n} =
\frac{S'(u(z)+\xi_i)}{S'(\xi_i)},
$$
we can reformulate the statement as $z$-independence 
of the ratio
$
\displaystyle{\frac{\p_{\lambda_j}Q(u(z), \xi_i )}{Q(u(z), \xi_j)-Q(u(z), \xi_i )}}.
$
The coefficients $\Gamma_{ij}$ are then found as
\beq\label{g2b}
\frac{\p_{\lambda_j}Q(u(z), \xi_i )}{Q(u(z), \xi_j)-Q(u(z), \xi_i )}=
\Gamma_{ij}
\eeq

\paragraph{The denominator $Q(u(z), \xi_j )-Q(u(z), \xi_i )$.}
The denominator of (\ref{g2b}) is
\beq\label{den0}
Q(u, \xi_j)-Q(u, \xi_i )=\frac{S'(u+\xi_j)S'(\xi_i)-S'(u+\xi_i)S'(\xi_j)}{S'(\xi_i)S'(\xi_j)},
\eeq
the numerator of which is expressed through theta functions thanks to (\ref{n2}) as follows:
$$
S'(u+\xi_j)S'(\xi_i)-S'(u+\xi_i)S'(\xi_j)
$$
$$
=\pi^2 \theta_4^4(0, \tau )
\frac{\theta_2(u+\xi_j)\theta_2(\xi_i)
\theta_1(u+\xi_i)\theta_1(\xi_j)-
\theta_1(u+\xi_j)\theta_1(\xi_i)
\theta_2(u+\xi_i)\theta_2(\xi_j)}{\theta_1(u+\xi_j)\theta_1(\xi_i)
\theta_1(u+\xi_i)\theta_1(\xi_j)}.
$$
Here and below in this appendix $\theta_a(x)=\theta_a(x, \frac{\tau}{2})$.
By subtracting (R9) in p.20 of \cite{mum:82} from (R8), we have a formula
\beq\label{den1}
\begin{array}{c}
\theta_2(x)\theta_2(y)\theta_1(u)\theta_1(v)-
\theta_1(x)\theta_1(y)\theta_2(u)\theta_2(v)
\\
=\, \theta_2(x_1)\theta_2(y_1)\theta_1(u_1)\theta_1(v_1)-
\theta_1(x_1)\theta_1(y_1)\theta_2(u_1)\theta_2(v_1).
\end{array}
\eeq
By setting $x\mapsto u+\xi_j$, $y\mapsto \xi_i$, $u\mapsto u+\xi_i$, $v\mapsto \xi_j$, the 
arguments in the right hand side become
$$
\begin{array}{lll}
x_1=\frac{1}{2}(x+y+u+v)& \phantom{aaaa} \mapsto \phantom{aaaa}&u+\xi_i+\xi_j,
\\
y_1=\frac{1}{2}(x+y-u-v)& \phantom{aaaa} \mapsto \phantom{aaaa}&0,
\\
u_1=\frac{1}{2}(x-y+u-v)& \phantom{aaaa} \mapsto \phantom{aaaa}&u,
\\
v_1=\frac{1}{2}(x-y-u+v)& \phantom{aaaa} \mapsto \phantom{aaaa}&\xi_j-\xi_i.
\end{array}
$$
Substituting (\ref{den1}) with this specialization, we have:
$$
S'(u+\xi_j)S'(\xi_i)-S'(u+\xi_i)S'(\xi_j)
=\, \pi^2 \theta_4^4(0, \tau )
\frac{\theta_2(u+\xi_i+\xi_j)\theta_2(0)\theta_1(u)\theta_1(\xi_j-\xi_i)}{\theta_1(u+\xi_j)\theta_1(\xi_i)
\theta_1(u+\xi_i)\theta_1(\xi_j)}.
$$
Thus the expression (\ref{den0}) is factorized as
\beq\label{den2}
Q(u, \xi_j)-Q(u, \xi_i )=\frac{\theta_2(u+\xi_i+\xi_j)
\theta_2(0)\theta_1(u)\theta_1(\xi_j-\xi_i)}{\theta_1(u+\xi_j)\theta_2(\xi_i)
\theta_1(u+\xi_i)\theta_2(\xi_j)}
\eeq
because of (\ref{n2}).
Passing to $\tau$ instead of $\frac{\tau}{2}$ in $u$-dependent factors with the help of 
the formulae
$$
\theta_1(x)\theta_2(0)=2\theta_1(x, \tau )\theta_4(x, \tau ), \qquad
\theta_2(x)\theta_2(0)=2\theta_2(x, \tau )\theta_3(x, \tau ),
$$
we finally obtain
\beq\label{den3}
\begin{array}{c}
Q(u, \xi_j)\! -\! Q(u, \xi_i )=
\displaystyle{\frac{\theta_2(u\! +\! \xi_i\! +\! \xi_j, \tau )\theta_3(u\! +\! \xi_i\! +\! \xi_j, \tau )
\theta_1(u, \tau )\theta_4(u, \tau )}{\theta_1(u+\xi_i, \tau )\theta_4(u+\xi_i, \tau )
\theta_1(u+\xi_j, \tau )\theta_4(u+\xi_j, \tau )}\,
\frac{\theta_1(\xi_j-\xi_i)\theta_2(0)}{\theta_2(\xi_i)\theta_2(\xi_j)}
}.
\end{array}
\eeq

\paragraph{The numerator $\p_{\lambda_j}Q(u(z), \xi_i )$.}
We compute the derivative honestly:
\beq\label{num1}
\frac{\p }{\p \lambda_j}\, \frac{S'(u+\xi_i)}{S'(\xi_i)}\, =\, 
\frac{1}{(S'(\xi_i))^2}\left (
\frac{\p S'(u+\xi_i)}{\p \lambda_j}\, S'(\xi_i)-\frac{\p S'(\xi_i)}{\p \lambda_j}\, S'(u+\xi_i)
\right ).
\eeq
Next we have to:
\begin{itemize}
\item use the chain rule and rewrite the expression using $S''$, $\dot S'$, 
$\displaystyle{\frac{\p u}{\p \lambda_j}}$, $\displaystyle{\frac{\p \xi_i}{\p \lambda_j}}$,
$\displaystyle{\frac{\p \tau}{\p \lambda_j}}$;
\item use the elliptic L\"owner equations (\ref{f1}) and the Gibbons-Tsarev system 
(\ref{GT1}), (\ref{GT2}) to represent the $\lambda_j$-derivatives of $u$, $\xi_i$ and $\tau$;
\item simplify the derivatives of $S$ in terms of theta functions.
\end{itemize}

The derivatives in the right hand side of (\ref{num1}) are:
$$
\frac{\p S'(u+\xi_i)}{\p \lambda_j} =S''(u+\xi_i)\left (\frac{\p u}{\p \lambda_j}+
\frac{\p \xi_i}{\p \lambda_j}\right ) +\dot S'(u+\xi_i)\frac{\p \tau}{\p \lambda_j},
$$
$$
\frac{\p S'(\xi_i)}{\p \lambda_j} =S''(\xi_i)
\frac{\p \xi_i}{\p \lambda_j} +\dot S'(\xi_i)\frac{\p \tau}{\p \lambda_j}.
$$
The elliptic L\"owner equation (\ref{f1}) and the Gibbons-Tsarev equation (\ref{GT1}) imply
$$
\frac{\p u}{\p \lambda_j}+
\frac{\p \xi_i}{\p \lambda_j}=\frac{1}{4\pi {\rm i}}\Bigl (
-\zeta_1(u+\xi_j)+\zeta_1(\xi_j-\xi_i)\Bigr ) \frac{\p \tau}{\p \lambda_j},
$$
where $\zeta_1(x)=\zeta_1(x, \frac{\tau}{2})$. Substituting (\ref{n5}) for $\dot S'$, we have:
\beq\label{num2}
\begin{array}{c}
\displaystyle{
4\pi {\rm i}\, (S'(\xi_i))^2\, \frac{\p }{\p \lambda_j}\, \frac{S'(u+\xi_i)}{S'(\xi_i)}}
\\ \\
=\, \displaystyle{\left [\phantom{\int}\!\!\!\! S''(u+\xi_i)S'(\xi_i)\Bigl (
-\zeta_1(u+\xi_j)+\zeta_1(\xi_j-\xi_i)+2\zeta_2(u+\xi_i, \tau )\Bigr )\right. }
\\ \\
-\, \displaystyle{ S''(\xi_i)S'(u+\xi_i)\Bigl (
-\zeta_1(\xi_j)+\zeta_1(\xi_j-\xi_i)+2\zeta_2(\xi_i, \tau )\Bigr ) }
\\ \\
+\, 2\displaystyle{\left.  S'(u+\xi_i)S'(\xi_i)\Bigl (-\wp_2 (u+\xi_i , \tau )+\wp_2 (\xi_i, \tau )\Bigr )
\phantom{\int}\!\!\!\!\right ]\frac{\p \tau}{\p \lambda_j}}.
\end{array}
\eeq

To proceed further, we need the identity
$$
-\zeta_1(x_1)+\zeta_1(x_2)+2\zeta_2(x_1-x_2, \tau )
$$
$$
=\,\pi \theta_2 (0, \tau)\theta_3(0, \tau )\theta_4^2(0, \tau )
\frac{\theta_1(x_1-x_2, \tau )\theta_4(x_1-x_2, \tau )\theta_2(x_1+x_2, \tau )}{\theta_1(x_1, \tau)
\theta_4(x_1, \tau)\theta_1(x_2, \tau )\theta_4(x_2, \tau)\theta_2(x_1-x_2, \tau )}
$$
(equation (A16) in \cite{akh-zab:14-1}). Using also (\ref{n11}) from 
Appendix A of the present paper, we represent (\ref{num2}) in the form
$$
\frac{4\pi {\rm i}\, (S'(\xi_i))^2}{\p \tau /\p \lambda_j}\, 
\frac{\p }{\p \lambda_j}\, \frac{S'(u+\xi_i)}{S'(\xi_i)}=
\pi \theta_2(0, \tau )\theta_3(0, \tau )\theta_4^2(0, \tau )\, f(u),
$$
where
$$
\begin{array}{lll}
f(u)&=&\displaystyle{S''(u+\xi_i)S'(\xi_i)
\frac{\theta_1(u+\xi_i,\tau)\theta_4(u+\xi_i,\tau)\theta_2(u+2\xi_j-\xi_i,\tau)}{\theta_1(u\! +\! \xi_j,\tau)
\theta_4(u\! +\! \xi_j,\tau)\theta_1(\xi_j\! -\! \xi_i,\tau)\theta_4(\xi_j\! -\! 
\xi_i,\tau)\theta_2(u\! +\! \xi_i,\tau)}
}
\\ && \\
&&-\, \displaystyle{ S''(\xi_i)S'(u+\xi_i)
\frac{\theta_1(\xi_i,\tau)\theta_4(\xi_i,\tau)\theta_2(2\xi_j-\xi_i,\tau)}{\theta_1(\xi_j,\tau)
\theta_4(\xi_j,\tau)\theta_1(\xi_j-\xi_i,\tau)\theta_2(\xi_i,\tau)}
}
\\ && \\
&&-\, 2\pi\displaystyle{S'(u+\xi_i)S'(\xi_i)\, \theta_2(0, \tau)\theta_3(0, \tau)
\frac{\theta_1(u,\tau)\theta_1(u+2\xi_i, \tau )}{\theta_2^2(\xi_i, \tau )\theta_2^2(u+\xi_i, \tau )}.
}
\end{array}
$$
Plugging here $S'$, $S''$ from (\ref{n2}), (\ref{n3}), we get:
\beq\label{num4}
\frac{4\pi {\rm i}\, (S'(\xi_i))^2}{\p \tau /\p \lambda_j}\, 
\frac{\p }{\p \lambda_j}\, \frac{S'(u+\xi_i)}{S'(\xi_i)}=
\pi^4 \theta_2^2(0, \tau )\theta_3^2(0, \tau )\theta_4^6(0, \tau )\, g(u),
\eeq
where the function $g(u)$ reads
{\small
$$
g(u)
$$
$$
=-\frac{\theta_4(2u\! +\! 2\xi_i, \tau)\theta_2(u\! +\! 2\xi_j \! -\! \xi_i,\tau)
\theta_2(\xi_i,\tau)\theta_3(\xi_i,\tau)
\theta_2(0, \tau )\theta_3(0, \tau )\theta_4(0, \tau )}{\theta_1(u\! +\! \xi_i,\tau)
\theta_4(u\! +\! \xi_i,\tau)\theta_1(u\! +\! \xi_j,\tau)\theta_4(u\! +\! \xi_j,\tau)
\theta_2(u\! +\! \xi_i,\tau)\theta_1(\xi_i,\tau)\theta_4(\xi_i,\tau)
\theta_1(\xi_j\! -\! \xi_i,\tau)\theta_4(\xi_j\! -\! \xi_i,\tau)}
$$
$$
+\, \frac{\theta_2(u+\xi_i,\tau)\theta_3(u+\xi_i,\tau)\theta_4(2\xi_i,\tau)
\theta_2(2\xi_j-\xi_i,\tau)
\theta_2(0, \tau )\theta_3(0, \tau )\theta_4(0, \tau )}{\theta_1(u\! +\! \xi_i,\tau)
\theta_4(u\! +\! \xi_i,\tau)\theta_1(\xi_i,\tau)\theta_4(\xi_i,\tau)
\theta_1(\xi_j,\tau)\theta_4(\xi_j,\tau)
\theta_1(\xi_j\! -\! \xi_i,\tau)\theta_4(\xi_j\! -\! \xi_i,\tau)\theta_2(\xi_i, \tau)}
$$
$$
-\, 2\,  \frac{\theta_1(u,\tau)\theta_1(u+2\xi_i,\tau)\theta_3(u+\xi_i,\tau)
\theta_3(\xi_i,\tau)}{\theta_1(u\! +\! \xi_i,\tau)
\theta_4(u\! +\! \xi_i,\tau)\theta_2(u+\xi_i,\tau)\theta_1(\xi_i,\tau)
\theta_4(\xi_i,\tau)\theta_2(\xi_i, \tau)}.
$$
}
It is easy to see that it is an elliptic function of $u$ with periods $1$, $\tau$ and four simple poles
in the fundamental parallelogram at the points $u=-\xi_i$, $u=-\xi_j$, $u=-\xi_i+\frac{\tau}{2}$,
$u=-\xi_j+\frac{\tau}{2}$. A possible pole at $u=-\xi_i+\frac{1}{2}$ cancels thanks to the identity
\beq\label{num3}
\theta_1(2x, \tau)\theta_2(0, \tau )\theta_3(0, \tau )\theta_4(0, \tau )=2
\theta_1(x,\tau)\theta_2(x,\tau)\theta_3(x,\tau)\theta_4(x,\tau).
\eeq
One can see that $g(u)$ has zeros at the four points $u=0$, $u=\frac{1}{2}-\xi_i-\xi_j$,
$u=\frac{\tau}{2}$ and $u=\frac{1+\tau}{2}-\xi_i-\xi_j$.
Therefore, $g(u)$ can be represented in the form
$$
g(u)=\frac{\theta_1(u,\tau)\theta_4(u,\tau)\theta_2(u+\xi_i+\xi_j,\tau)
\theta_3(u+\xi_i+\xi_j,\tau)}{\theta_1(u\! +\! \xi_i,\tau)
\theta_4(u\! +\! \xi_i,\tau)\theta_1(u\! +\! \xi_j,\tau)
\theta_4(u\! +\! \xi_j,\tau)}
$$
$$
\times \, \frac{C}{
\theta_1(\xi_i,\tau)\theta_4(\xi_i,\tau)\theta_1(\xi_j,\tau)\theta_4(\xi_j,\tau)
\theta_1(\xi_j\! -\! \xi_i,\tau)\theta_4(\xi_j\! -\! \xi_i,\tau)}
$$
with some constant $C$. The constant can be easily found by tending $u\to -\xi_j$:
$$
C= \theta_2(0, \tau )\theta_3(0, \tau )\theta_4(0, \tau )\theta_4(2\xi_i-2\xi_j,\tau).
$$ 

\paragraph{Calculation of $\Gamma_{ij}$.}
Dividing (\ref{num4}) by (\ref{den3}) we see that the $u$-dependent factors cancel
(and, therefore, the $z$-dependence disappears)
and we are left with
$$
\frac{4\pi {\rm i}\, (S'(\xi_i))^2}{\p \tau /\p \lambda_j}\,\Gamma_{ij}=
\frac{8\pi^4 \, \theta_2^3(0, \tau )\theta_3^3(0, \tau )\theta_4^7(0, \tau )}{\theta_2^4(0, \frac{\tau}{2})}
\, \frac{\theta_4(2\xi_i-2\xi_j,\tau)\theta_2(\xi_i,\frac{\tau}{2})
\theta_2(\xi_j,\frac{\tau}{2})}{\theta_1(\xi_i,\frac{\tau}{2})
\theta_1(\xi_j,\frac{\tau}{2})\theta_1^2(\xi_j-\xi_i,\frac{\tau}{2})}.
$$
Using (\ref{n2}), (\ref{n3}) and some identities for theta-constants, we obtain
the final result for $\Gamma_{ij}$:
\beq\label{g3b}
\Gamma_{ij}=-\, \frac{1}{4\pi {\rm i}}\, \frac{S'(\xi_j)}{S'(\xi_i)}
\, S''(\xi_i-\xi_j)\, \frac{\p \tau}{\p \lambda_j}.
\eeq

\subsection*{Appendix C: proof of $\Gamma_{ij}=\frac{1}{2}\, \p_{\lambda_j}\log g_i$}
\def\theequation{C\arabic{equation}}
\setcounter{equation}{0}

Here we prove that 
\beq\label{c1a}
\Gamma_{ij}=-\, \frac{1}{4\pi {\rm i}}\, \frac{S'(\xi_j)}{S'(\xi_i)}
\, S''(\xi_i-\xi_j)\, \frac{\p \tau}{\p \lambda_j}=\frac{1}{2}\, \p_{\lambda_j}\log g_i,
\eeq
where
$$
g_i =\frac{1}{4\pi {\rm i}}\, (S'(\xi_i))^2 \frac{\p \tau}{\p \lambda_i}.
$$

Let us take the derivative:
$$
\frac{\p }{\p \lambda_j}\left (\log S'(\xi_i)+\frac{1}{2}\log \frac{\p \tau}{\p \lambda_i}\right )
=\frac{S''(\xi_i)}{S'(\xi_i)}\, \frac{\p \xi_i}{\p \lambda_j}+
\frac{\dot S'(\xi_i)}{S'(\xi_i)}\, \frac{\p \tau}{\p \lambda_j}+
\frac{1}{2(\p \tau /\p \lambda_i)}\, \frac{\p^2 \tau}{\p \lambda_i \p \lambda_j}.
$$
Using the elliptic Gibbons-Tsarev system
\beq\label{GT1a}
\frac{\p \xi_i}{\p \lambda_j}=\frac{1}{4\pi {\rm i}}\, \Bigl (
\zeta_1 (-\xi_i+\xi_j, \tau ')-\zeta_1(\xi_j, \tau ')\Bigr )\, \frac{\p \tau}{\p \lambda_j}
\eeq
\beq\label{GT2a}
\frac{\p^2 \tau}{\p \lambda_i \p \lambda_j}=\frac{1}{2\pi {\rm i}}\,
\wp _1(\xi_i-\xi_j, \tau ')\frac{\p \tau}{\p \lambda_i}\frac{\p \tau}{\p \lambda_j}
\eeq
and the formula (\ref{n5}) for $\dot S'$, we get
$$
\frac{1}{2}\, \frac{\p }{\p \lambda_j}\log g_i =\frac{1}{4\pi {\rm i}S'(\xi_i)}\, \frac{\p \tau}{\p \lambda_j}
\Bigl  (  S''(\xi_i)(\zeta_1 (\xi_j -\xi_i, \tau')-\zeta_1 (\xi_j, \tau'))  
$$
$$
+\, S'(\xi_i)(\wp_1(\xi_i-\xi_j, \tau')-2\wp_2(\xi_i, \tau ))+2S''(\xi_i)\zeta_2(\xi_i, \tau )
\Bigr ).
$$
Comparing with (\ref{c1a}), we see that we should prove the identity
\beq\label{c5}
\begin{array}{c}
S''(\xi_i)\zeta_1 (\xi_j -\xi_i, \tau')-S''(\xi_i)\zeta_1(\xi_j, \tau')
 +S'(\xi_i)\wp_1(\xi_i-\xi_j, \tau')
\\ \\
+\,\, 2S''(\xi_i)\zeta_2(\xi_i, \tau )-2S'(\xi_i)\wp_2(\xi_i, \tau )+S'(\xi_j)
S''(\xi_j-\xi_i)\, =0.
\end{array}
\eeq
The way to prove it is standard. It is easy to see that
as a function of $\xi_j$ the left hand side is an elliptic function with periods
$1$, $\tau'$. It may have singularities at the points $\xi_j=\xi_i$ and $\xi_j=0$ only.
Setting $\xi_j =\xi_i +\varepsilon$ and $\xi_j =\varepsilon$, one can see that the 
principal parts as $\varepsilon \to 0$ (double and simple poles) cancel, so the left hand side 
is a regular function and thus it is constant in $\xi_j$. 

To find the constant, let us evaluate
this expression at the regular point $\xi_j=\xi_i +\frac{1}{2}$. We have, using (\ref{n9}), (\ref{n10}):
$$
\begin{array}{l}
S''(\xi_i)\Bigl  (2\zeta_2(\xi_i, \tau )-\zeta_2(\xi_i, \tau ')\Bigr )
-2S'(\xi_i)\wp_2(\xi_i, \tau )+S''(\frac{1}{2})S'(\xi_i +\frac{1}{2})
+\wp_1(\frac{1}{2}, \tau ')S'(\xi_i)
\end{array}
$$
$$
\begin{array}{l}
=\underbrace{S''(\xi_i)S'(\xi_i +\frac{1}{2})+S'(\xi_i)S''(\xi_i +
\frac{1}{2})}_{=0 \; \mbox{\scriptsize{by virtue of (\ref{n6a}})}}
-S'(\xi_i)\Bigl (\wp_2(\xi_i, \tau')-\wp_2(0, \tau')\Bigr )
+S''(\frac{1}{2})S'(\xi_i +\frac{1}{2})
\end{array}
$$
\beq\label{c6}
=\,\, -S'(\xi_i)\Bigl (\wp_2(\xi_i, \tau')-\wp_2(0, \tau')\Bigr )
+S''(\frac{1}{2})S'(\xi_i +\frac{1}{2}).
\eeq
Now, using (\ref{n2}), (\ref{n3}) and (\ref{n11}) we conclude that this is equal to $0$.

\subsection*{Appendix D: proof of $\p_{\lambda_k}\Gamma_{ij}=
\Gamma_{ij}\Gamma_{jk}+\Gamma_{ik}\Gamma_{kj}-\Gamma_{ik}\Gamma_{ij}$}
\def\theequation{D\arabic{equation}}
\setcounter{equation}{0}

Taking the derivative $\p_{\lambda_k}\Gamma_{ij}$ ($\Gamma_{ij}$ 
is given by (\ref{c1a})) with the help of the Gibbons-Tsarev 
system and equation (\ref{n5}), we get
$$
\p_{\lambda_k}\Gamma_{ij}=-\frac{1}{(4\pi {\rm i})^2 S'(\xi_i)}\,
\frac{\p \tau}{\p \lambda_j}\, \frac{\p \tau}{\p \lambda_k}
$$
$$
\times \, \left [\phantom{\int}\!\!\!\! S'''(\xi_i -\xi_j)S'(\xi_j)
\Bigl (\zeta_1(\xi_k-\xi_i, \tau ')-\zeta_1(\xi_k-\xi_j, \tau ')+2\zeta_2(\xi_i-\xi_j, \tau)
\Bigr )\right.
$$
$$
-4S''(\xi_i -\xi_j)S'(\xi_j)\wp_2 (\xi_i-\xi_j, \tau )-2
S'(\xi_i -\xi_j)S'(\xi_j)\wp_2' (\xi_i-\xi_j, \tau )
$$
$$
+\, 2 S''(\xi_i -\xi_j)S'(\xi_j)\wp_1 (\xi_j-\xi_k, \tau ')
$$
$$
+\, S''(\xi_i -\xi_j)S''(\xi_j)\Bigl (-\zeta_1(\xi_k, \tau ')+\zeta_1(\xi_k-\xi_j, 
\tau ')+2\zeta_2(\xi_j, \tau)
\Bigr )
$$
$$
-\, S''(\xi_i -\xi_j)S''(\xi_i)\frac{S'(\xi_j)}{S'(\xi_i)}
\Bigl (-\zeta_1(\xi_k, \tau ')+\zeta_1(\xi_k-\xi_i, 
\tau ')+2\zeta_2(\xi_i, \tau)
\Bigr )
$$
$$
\left. +\, 2S''(\xi_i -\xi_j)S'(\xi_j)\Bigl (\wp_2(\xi_i, \tau )-\wp_2(\xi_j, \tau )\Bigr )
\phantom{\int}\!\!\!\! \right ].
$$
We recall that $\tau ' =\frac{\tau}{2}$.
The functions $\zeta_2 (x, \tau)$, $\wp_2(x, \tau )$ and $\wp_2'(x, \tau )$
can be transformed to the functions $\zeta_2 (x, \tau ')$, $\wp_2(x, \tau ' )$ and $\wp_2'(x, \tau ' )$
with the help of (\ref{n9}), (\ref{n10}) (and derivative of (\ref{n10})). 
The expressions containing products of derivatives of the $S$-function then cancel by
virtue of (\ref{n6a}) (one should take derivatives of this equation). In this way we get
\beq\label{gg1}
-\p_{\lambda_k}\Gamma_{ij}-\Gamma_{ik}\Gamma_{ij}+
\Gamma_{ij}\Gamma_{jk}+\Gamma_{ik}\Gamma_{kj}=
\frac{S'(\xi_j)}{(4\pi {\rm i})^2 S'(\xi_i)}\,
\frac{\p \tau}{\p \lambda_j}\, \frac{\p \tau}{\p \lambda_k}\, h(\xi_i, \xi_j, \xi_k),
\eeq
where
$$
h=S'''(\xi_i -\xi_j)
\Bigl (\zeta_1(\xi_k-\xi_i)-\zeta_1(\xi_k-\xi_j)+\zeta_2(\xi_i-\xi_j)
\Bigr )
$$
$$
-2S''(\xi_i -\xi_j)\wp_2 (\xi_i-\xi_j )-
S'(\xi_i -\xi_j)\wp_2' (\xi_i-\xi_j)
$$
$$
+\, 2 S''(\xi_i -\xi_j)\wp_1 (\xi_j-\xi_k)\phantom{aaaaaaaaaaaaa}
$$
$$
+\, S''(\xi_i -\xi_j)\frac{S''(\xi_j)}{S'(\xi_j)}
\Bigl (-\zeta_1(\xi_k)+\zeta_1(\xi_k-\xi_j)+\zeta_2(\xi_j)
\Bigr )
$$
$$
-\, S''(\xi_i -\xi_j)\frac{S''(\xi_i)}{S'(\xi_i)}
\Bigl (-\zeta_1(\xi_k)+\zeta_1(\xi_k-\xi_i)+\zeta_2(\xi_i)
\Bigr )
$$
$$
+\, S''(\xi_i -\xi_j)\Bigl (\wp_2(\xi_i)-\wp_2(\xi_j )\Bigr ) \phantom{aaaaaaaaaaaaa}
$$
$$
-\, \frac{S'(\xi_k)}{S'(\xi_i)}\, S''(\xi_k-\xi_i)S''(\xi_i-\xi_j)+
\frac{S'(\xi_k)}{S'(\xi_j)}\, S''(\xi_k-\xi_j)S''(\xi_i-\xi_j)
$$
$$
+\, S''(\xi_k-\xi_i)S''(\xi_k-\xi_j). \phantom{aaaaaaaaaaaaa}
$$
Here all $\zeta$- and $\wp$-functions depend on $\tau '$. As a function of 
$\xi_k$, $h$ is a double-periodic function with periods $1$, $\tau '$ and 
possible poles at the points $\xi_k=\xi_i$, $\xi_k =\xi_j$ and $\xi_k=0$
in the fundamental parallelogram. 
One can see that the principal parts of the expansions around these points vanish,
so the function is regular and thus it is a constant in $\xi_k$. To find the constant,
we evaluate $h$ at the point $\xi_k=\frac{1}{2}$:
$$
\begin{array}{lll}
h(\xi_i, \xi_j, \frac{1}{2})&=&
S'''(\xi_i -\xi_j)
\Bigl (-\zeta_2(\xi_i)+\zeta_2(\xi_j)+\zeta_2(\xi_i-\xi_j)
\Bigr )
\\ &&\\
&-&
2S''(\xi_i -\xi_j)\wp_2 (\xi_i-\xi_j )-
S'(\xi_i -\xi_j)\wp_2' (\xi_i-\xi_j)
\\&&\\
&+&S''(\xi_i-\xi_j)\wp_2(\xi_i)+S''(\xi_i-\xi_j)\wp_2(\xi_j)
+S''(\xi_i +\frac{1}{2})S''(\xi_j +\frac{1}{2}).
\end{array}
$$
We want to show that $h(\xi_i, \xi_j, \frac{1}{2})=0$, then 
$h(\xi_i, \xi_j, \xi_k)=0$ and we are done. Let us denote
$H(\xi_i)=h(\xi_i, \xi_j, \frac{1}{2})$. The function $H$ has the following 
quasiperiodicity properties: $H(\xi_i+1)=H(\xi_i)$, 
$H(\xi_i+\tau ')=-H(\xi_i)$. It is enough to show that $H(\xi_i)$ is regular
at the possible singular points $\xi_i=\xi_j$, $\xi_i=\xi_j+\frac{1}{2}$,
$\xi_i=\frac{1}{2}$ and $H(0)=0$. The regularity is checked by a direct inspection.
At $\xi_i=0$ we have:
$$
\begin{array}{l}
H(0)=-S'(\xi_j)\wp_2'(\xi_j0+S''(\xi_j)\wp_2(0)-S''(\xi_j)\wp_2(\xi_j)
+S''(\frac{1}{2})S''(\xi_j+\frac{1}{2})
\end{array}
$$
which is the derivative of (\ref{c6}) and thus equal to zero.

\subsection*{Appendix E: proof of equation (\ref{cq3})}
\def\theequation{E\arabic{equation}}
\setcounter{equation}{0}

In order to prove (\ref{cq3}) we start from (\ref{f202}):
$$
4\pi {\rm i}\p_{\lambda_j}S(u)=S'(\xi_j)S'(u+\xi_j)\frac{\p \tau}{\p \lambda_j}.
$$
To take another derivative we need to know
$\p_{\lambda_i}S'(u+\xi_j)$ and $\p_{\lambda_i}S'(\xi_j)$
which are calculated using the elliptic L\"owner equation,
the Gibbons-Tsarev equations,
equations (\ref{n5}), (\ref{n9}), (\ref{n10}) and the derivative of equation (\ref{n6a}):
$$
4\pi {\rm i}\p_{\lambda_i}S'(u+\xi_j)=
\Bigl [ S''(u+\xi_j)\Bigl ( -\zeta_1(u+\xi_i)+\zeta_1(\xi_i-\xi_j)+\zeta_2(u+\xi_j)\Bigr )
-S'(u+\xi_j)\wp_2(u+\xi_j)\Bigr ]\frac{\p \tau}{\p \lambda_i}.
$$
Here and below all $\zeta$- and $\wp$-functions have modular parameter $\frac{\tau}{2}$. 
At $z=\infty$ ($u=0$) we get from here
$$
4\pi {\rm i}\p_{\lambda_i}S'(\xi_j)=
\Bigl [ S''(\xi_j)\Bigl ( -\zeta_1(\xi_i)+\zeta_1(\xi_i-\xi_j)+\zeta_2(\xi_j)\Bigr )
-S'(\xi_j)\wp_2(\xi_j)\Bigr ]\frac{\p \tau}{\p \lambda_i}.
$$
Combining all necessary equations together, we have:
$$
\p_{\lambda_i}\p_{\lambda_j}S(u)-\Gamma_{ij}\p_{\lambda_i}S(u)-
\Gamma_{ji}\p_{\lambda_j}S(u)
=\frac{1}{(4\pi {\rm i})^2}\, \frac{\p \tau}{\p \lambda_i}\, \frac{\p \tau}{\p \lambda_j}\, f(\xi_i, \xi_j, u),
$$
where
$$
f(\xi_i, \xi_j, u)= 2S'(\xi_j)S'(u+\xi_j)\wp_1(\xi_i-\xi_j)
$$
$$
+\, S'(\xi_j)S''(u+\xi_j)\Bigl (-\zeta_1(u+\xi_i)+\zeta_1(\xi_i-\xi_j)+\zeta_2(u+\xi_j)\Bigr )
-S'(\xi_j)S'(u+\xi_j)\wp_2(u+\xi_j)
$$
$$
+\, S''(\xi_j)S'(u+\xi_j)\Bigl (-\zeta_1(\xi_i)+\zeta_1(\xi_i-\xi_j)+\zeta_2(\xi_j)\Bigr )
-S'(\xi_j)S'(u+\xi_j)\wp_2(\xi_j)
$$
$$
+\, S''(\xi_i-\xi_j)\Bigl (S'(\xi_i)S'(u+\xi_j)+S'(\xi_j)S'(u+\xi_i)\Bigr ).
$$
We want to prove that $f(\xi_i, \xi_j, u)=0$. The argument is standard.
As a function of $\xi_i$ $f(\xi_i, \xi_j, u)$ is an elliptic function with periods 
$1$ and $\frac{\tau}{2}$ and possible poles at the points $\xi_i=\xi_j$, $\xi_i=-u$
and $\xi_i=0$. Expanding around these points, one can see that the principal parts vanish,
so the function is constant in $\xi_i$. To find the constant, we evaluate it at the regular point
$\xi_i=\xi_j+\frac{1}{2}$. We have:
$$
\begin{array}{c}
f(\xi_j+\frac{1}{2}, \xi_j, u)=2S'(\xi_j)S'(u+\xi_j)\wp_2(0)
\\ \\
-
S'(\xi_j)S'(u+\xi_j)\wp_2(u+\xi_j)-
S'(\xi_j)S'(u+\xi_j)\wp_2(\xi_j)
\\ \\
+S''(\frac{1}{2})\Bigl ( S'(\xi_j+\frac{1}{2})S'(u+\xi_j)+S'(\xi_j)
S'(u+\xi_j+\frac{1}{2})\Bigr )
\\ \\
=\, S'(\xi_j)S'(u+\xi_j)\displaystyle{
\left [2\wp_2(0)-\wp_2(u+\xi_j)-\wp_2(\xi_j)\phantom{\int_A^B}\right.}
\\ \\
\displaystyle{\left. +\, S''(\frac{1}{2})
\left (\frac{S'(\xi_j+\frac{1}{2})}{S'(\xi_j)}+\frac{S'(u+\xi_j+\frac{1}{2})}{S'(u+\xi_j)}
\right ) \right ].
}
\end{array}
$$
Using (\ref{n11}), (\ref{n2}) and (\ref{n3}) (from which we find 
$S''(\frac{1}{2})=-\pi^2 \theta_3^2(0, \frac{\tau}{2})\theta_4^2(0, \frac{\tau}{2})$), one can see that
the expression in the square brackets equals zero, so $f=0$. 

\section*{Acknowledgments}

The work of V.A. was supported in part by the RFBR grant 16-01-00562.
The work of T.T. was partly prepared within the framework of the Academic Fund Program at the 
National Research University Higher School of Economics (HSE) in 2015--2016 (grant ¹15-01-0102)  
and supported within the framework of a subsidy granted to the HSE by the Government 
of the Russian Federation for the implementation of the Global Competitiveness Program.
The work of A.Z. was
supported by RSF grant 16-11-10160.

\end{document}